\documentclass[a4paper]{spie}  

\usepackage{amsmath,amsfonts,amssymb}
\usepackage{graphicx}
\usepackage[colorlinks=true, allcolors=blue]{hyperref}

\title{What is your favorite transient event? \\SOXS is almost ready to observe!}

\author[a,*]{Kalyan Kumar Radhakrishnan Santhakumari}
\author[a, b]{Federico Battaini}
\author[a]{Simone Di Filippo}
\author[a, b]{Silvio Di Rosa}
\author[c]{Lorenzo Cabona}
\author[a]{Riccardo Claudi}
\author[a]{Luigi Lessio}
\author[a]{Marco Dima}
\author[c]{David Young}
\author[d]{Marco Landoni}
\author[e]{Mirko Colapietro}
\author[e]{Sergio D'Orsi}
\author[d]{Matteo Aliverti}
\author[d]{Matteo Genoni}
\author[f]{Matteo Munari}
\author[e]{Ricardo Zanmar S\'anchez}
\author[l]{Fabrizio Vitali}
\author[a]{Davide Ricci}
\author[e]{Pietro Schipani}
\author[d]{Sergio Campana}
\author[g]{Jani Achr\'en}
\author[h]{Jos\'e Araiza-Dur\'an}
\author[i]{Iair Arcavi}
\author[a]{Andrea Baruffolo}
\author[j]{Sagi Ben-Ami}
\author[j]{Alex Bitchkovsky}
\author[h]{Anna Brucalassi}
\author[j]{Rachel Bruch}
\author[e]{Giulio Capasso}
\author[a]{Enrico Cappellaro}
\author[k]{Rosario Cosentino}
\author[l]{Francesco D'Alessio}
\author[d]{Paolo D'Avanzo}
\author[e]{Massimo Della Valle}
\author[f]{Rosario Di Benedetto}
\author[j]{Avishay Gal-Yam}
\author[k]{Marcos Hernandez Diaz}
\author[j]{Ofir Hershko}
\author[m,n]{Jari Kotilainen}
\author[m,n]{Hanindyo Kuncarayakti}
\author[o]{Gianluca Li Causi}
\author[a]{Luca Marafatto}
\author[f]{Eugenio Martinetti}
\author[e]{Laurent Marty}
\author[n]{Seppo Mattila}
\author[f]{Antonio Miccich\`e}
\author[f]{Gaetano Nicotra}
\author[d]{Luca Oggioni}
\author[k]{Hector Perez Ventura}
\author[d]{Giorgio Pariani}
\author[p]{Giuliano Pignata}
\author[j]{Michael Rappaport}
\author[d]{Marco Riva}
\author[q]{Adam Rubin}
\author[a]{Bernardo Salasnich}
\author[e]{Salvatore Savarese}
\author[r,f]{Salvatore Scuderi}
\author[c]{Steven Smartt}
\author[s]{Maximilian Stritzinger}

\affil[a]{INAF – Osservatorio Astronomico di Padova, Vicolo dell’Osservatorio 5, I-35122, Padua, Italy}
\affil[b]{Dipartimento di Fisica e Astronomia ``G. Galilei'', Universit\`a di Padova, Italy }
\affil[c]{Astrophysics Research Centre, Queen’s University Belfast, Belfast, BT7 1NN, UK}
\affil[d]{INAF – Osservatorio Astronomico di Brera, Via Bianchi 46, I-23807, Merate, Italy}
\affil[e]{INAF – Osservatorio Astronomico di Capodimonte, Sal. Moiariello 16, I-80131, Naples, Italy}
\affil[f]{INAF – Osservatorio Astrofisico di Catania, Via S. Sofia 78 30, I-95123 Catania, Italy}
\affil[g]{Incident Angle Oy, Capsiankatu 4 A 29, FI-20320 Turku, Finland}
\affil[h]{INAF-Osservatorio Astrofisico di Arcetri, Largo Enrico Fermi, 5, I-50125 Firenze, Italy}
\affil[i]{The School of Physics and Astronomy, Tel Aviv University, Tel Aviv 69978, Israel}
\affil[j]{Weizmann Institute of Science, Herzl St 234, Rehovot, 7610001, Israel}
\affil[k]{FGG-INAF, TNG, Rambla J.A. Fern\'andez P\'erez 7, E-38712 Bren\~a Baja (TF), Spain}
\affil[l]{INAF – Osservatorio Astronomico di Roma, Via Frascati 33, I-00078 M. Porzio Catone, Italy} 
\affil[m]{Finnish Centre for Astronomy with ESO (FINCA), FI-20014 University of Turku, Finland} 
\affil[n]{Tuorla Observatory, Dept. of Physics and Astronomy, FI-20014 University of Turku, Finland} 
\affil[o]{INAF - Istituto di Astrofisica e Planetologia Spaziali, Rome, Italy}
\affil[p]{Instituto de Alta Investigaci\'on, Universidad de Tarapac\'a, Arica, Casilla 7D, Chile}

\affil[q]{ESO, Karl Schwarzschild Strasse 2, D-85748, Garching bei M\"unchen, Germany} 
\affil[r]{INAF – Istituto di Astrofisica Spaziale e Fisica Cosmica, Milan, Via Corti 12, I-20133, Italy}
\affil[s]{Aarhus University, Ny Munkegade 120, D-8000 Aarhus, Denmark}

\authorinfo{*KKRS: kalyan[dot]radhakrishnan[at]inaf[dot]it}

\begin{document} 
\maketitle

\begin{abstract}

The Son Of X-Shooter (SOXS) will be the specialized facility to observe any transient event with a flexible scheduler at the ESO New Technology Telescope (NTT) at La Silla, Chile. SOXS is a single object spectrograph offering simultaneous spectral coverage in UV-VIS (350-850 nm) and NIR (800-2000 nm) wavelength regimes with an average of R$\sim$4500 for a 1” slit. SOXS also has imaging capabilities in the visible wavelength regime. Currently, SOXS is being integrated at the INAF-Astronomical Observatory of Padova. Subsystem- and system-level tests and verification are ongoing to ensure and confirm that every requirement and performance are met. In this paper, we report on the integration and verification of SOXS as the team and the instrument prepare for the Preliminary Acceptance Europe (PAE).
\end{abstract}


\section{INTRODUCTION}
\label{sec:intro}  

If you have a favorite transient event to be observed, get ready! The SOXS instrument is almost there\cite{pietro2024,pasquini2024}. The system-level integration and verification of this new, specialized, facility instrument to observe any transient event in the UV, VIS, and NIR wavelengths is in the final stages at the INAF-Osservatorio Astronomico di Padova (INAF-OAPD) laboratory in Padova, Italy\cite{aliverti-2022,pietro-2022}. The visits related to the Preliminary Acceptance Europe (PAE) review of the instrument have already started.

SOXS will be mounted on one of the Nasmyth platforms of the 3.58m ESO New Technology Telescope (NTT) at the La Silla Observatory in Chile. To facilitate maximum usage of the instrument, SOXS has adopted to work with a flexible scheduler\cite{laura2024,laura-2022}. This single object spectrograph offers simultaneous spectral coverage in UV-VIS (350-850 nm) and NIR (800-2000 nm) wavelength regimes with an average of R$\sim$4500 for a 1” slit. SOXS also has imaging and photometric capabilities in the visible wavelength (360-970 nm) regime.

The SOXS consortium has a relatively large geographic spread; therefore, the Assembly Integration and Verification(AIV)of SOXS followed a modular approach \cite{riccardo2024,riccardo-2022}. Each of the five main subsystems of SOXS, namely the Common Path (CP), the Calibration Unit (CBX), the Acquisition Camera (AC), the UVVIS Spectrograph, and the NIR Spectrograph, underwent internal alignment and testing in the respective consortium institutes and delivered to INAF-OAPD, where the subsystems were integrated, tested stand-alone, and together with other subsystems, including the SOXS electronics and software.
c
\begin{figure} [ht]
\begin{center}
\begin{tabular}{c} 
\includegraphics[width=12cm]{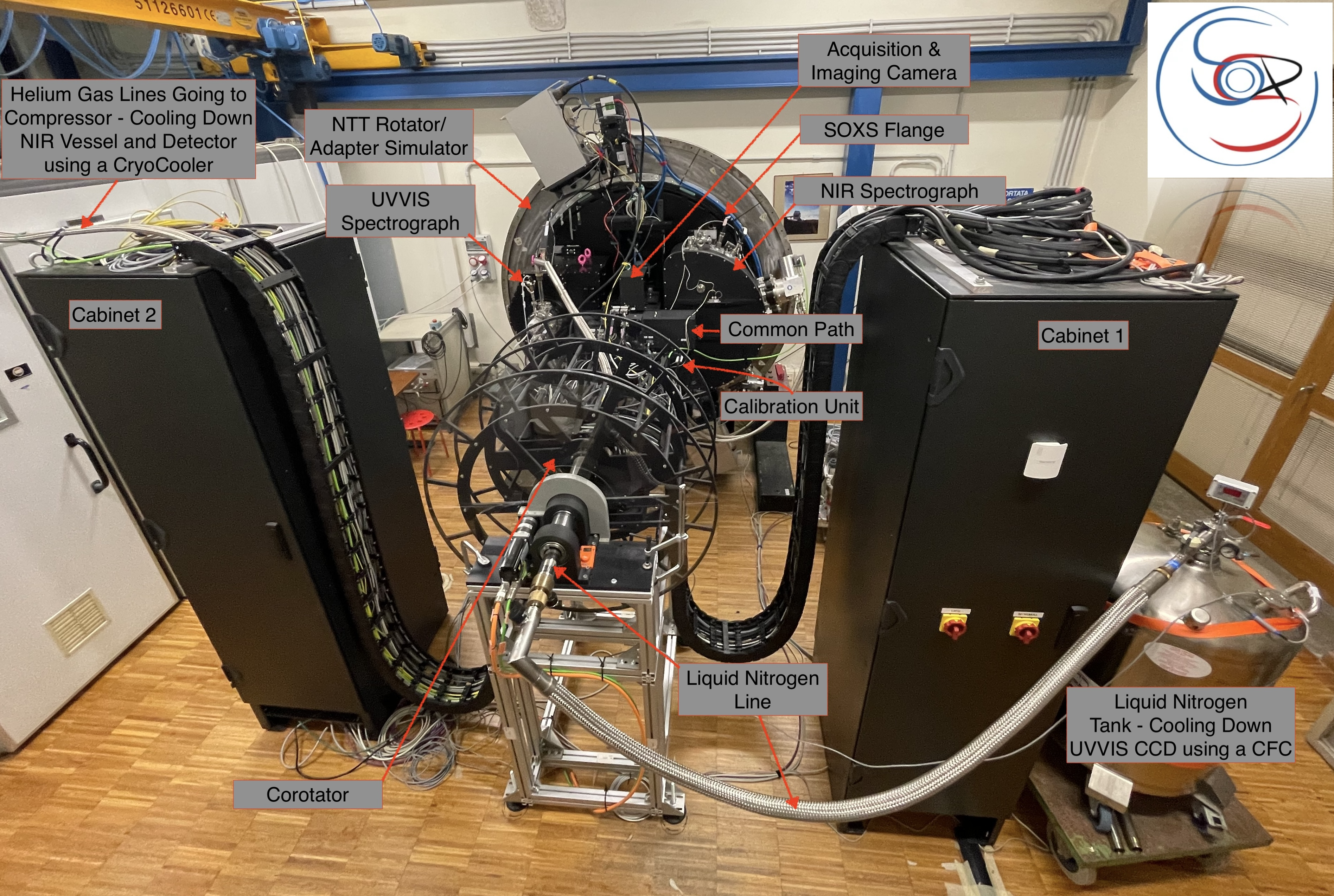}
\end{tabular}
\end{center}
\caption[example] 
   { \label{fig:SOXS-FULL} The full integrated SOXS system at the INAF-OAPD laboratory.}
\end{figure} 

System-level testing and verification are ongoing to ensure and confirm that every requirement and performance are met. In addition, tests are carried out related to the cryo-vacuum systems (for the UVVIS and NIR spectrographs), the electronics, water-cooling systems, instrument control software, template and OBs’, alarm systems, and the data reduction pipeline. Figure \ref{fig:SOXS-FULL} shows the fully integrated SOXS instrument at INAF-OAPD. In this article, we report on the integration and verification of SOXS as the team and the instrument prepare for the PAE.

\section{SOXS and its main features}

SOXS, the successor to the SOFI instrument at the Nasmyth platform of the NTT, is structurally supported by the SOXS Flange. This flange, which weighs about 355 kg, including the support structure, is directly mounted on the NTT derotator structure, providing a stable foundation for the instrument. 

\begin{figure} [ht]
\begin{center}
\begin{tabular}{c} 
\includegraphics[width=14cm]{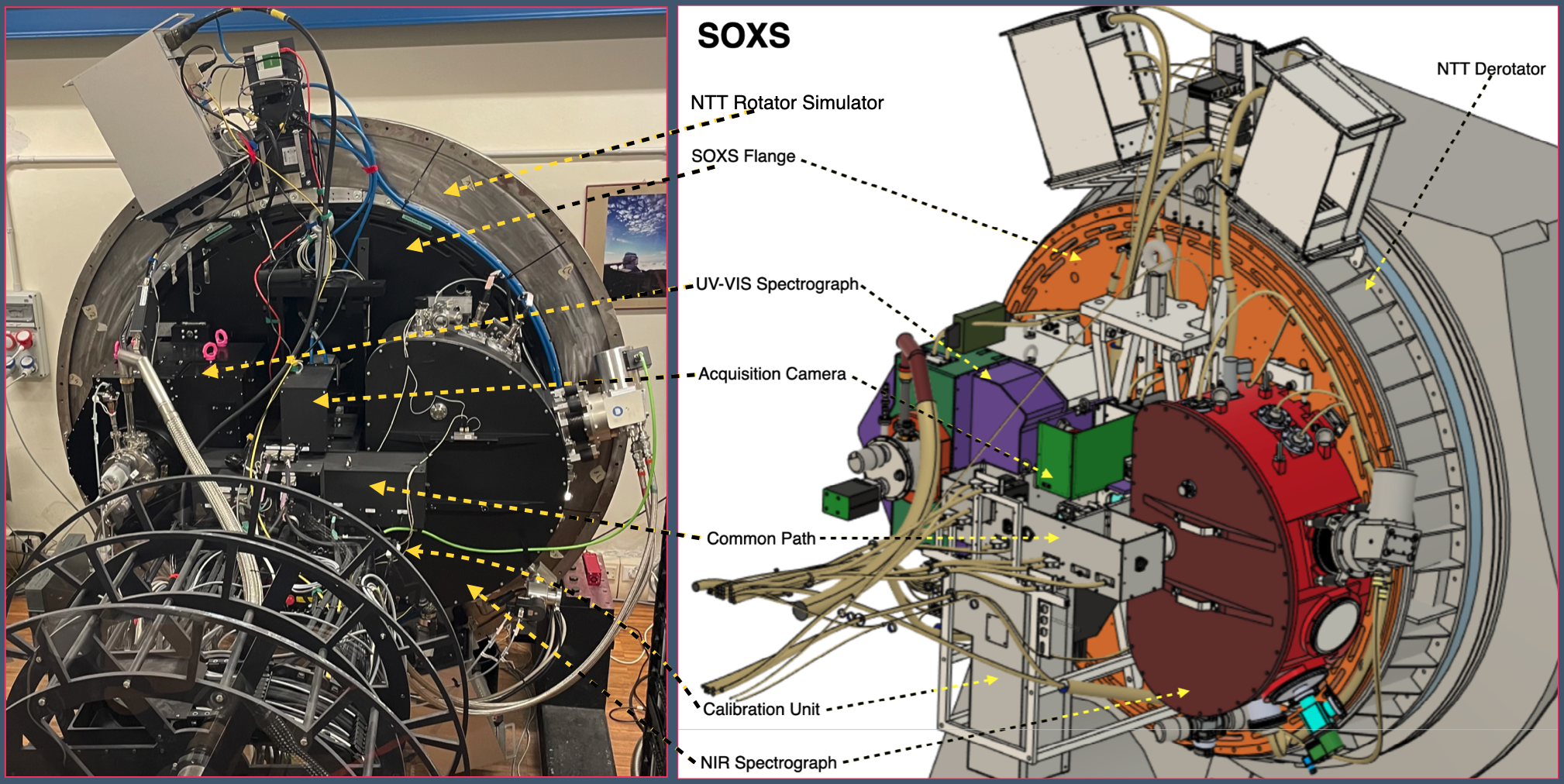}
\end{tabular}
\end{center}
\caption[example] 
   { \label{fig:SOXS-RvC} Fully integrated SOXS with its CAD image.}
\end{figure}

The rest of the SOXS are mounted on the flange except for UVVIS NGC\footnote{NGC stands for New Generation Control which facilitates the detector control}, NIR NGC, and the power supply and water cooling line support structure, which are also mounted directly on the NTT derotator (see Figure \ref{fig:SOXS-RvC}).

The three sub-systems - CP, UVVIS spectrograph, and NIR spectrograph are directly mounted on the SOXS flange using three (sphere, cylinder, and washer) Kinematic Mounts (KMs). The positioning repeatability of the sub-system using the KMs is measured to be within 100 $\mu$m. The CBX and AC are mounted on the CP using dedicated KMs. There are also the cryo-vacuum system, electronics cabinets, corotator system, instrument control software, data reduction pipeline, and water-cooling system. These auxiliary systems must work together to make the SOXS work to its full potential. In addition to the system-level functional and opto-mechanical performance testing, all the auxiliary systems were tested and optimized in Padova. Before going into details of each of these activities, the main SOXS scientific capabilities are tabulated in Table \ref{tab:SOXS-numbers}.
\begin{table}[ht]
\caption{SOXS scientific characteristics.} 
\label{tab:SOXS-numbers}
\begin{center}       
\begin{tabular}{|l|l|l|l|} 
\hline
\rule[-1ex]{0pt}{3.5ex}  \textbf{What} & \textbf{UVVIS} & \textbf{NIR} & \textbf{AC}  \\
\hline
\rule[-1ex]{0pt}{3.5ex}  \textbf{Spectral Range / Filters} & 350-850 nm & 800-2000 nm & u, g, r, i, z, Y \& V  \\
\hline
\rule[-1ex]{0pt}{3.5ex}  \textbf{Resolution (1" slit)} & $\sim$ 4500 & $\sim$5000& - \\
\hline
\rule[-1ex]{0pt}{3.5ex}  \textbf{Slit Widths} & 0.5" ,1", 1.5", 5" & 0.5" ,1", 1.5", 5" & FoV = 3.5' x 3.5' \\
\hline
\rule[-1ex]{0pt}{3.5ex}  \textbf{Slit Height} & 12" & 12" & - \\
\hline 
\rule[-1ex]{0pt}{3.5ex}  \textbf{Detector}  & e2V CCD44-82 2k x4k & Teledyne H2RG 2k x 2k & Andor iKon M-934 1k x 1k\\
\hline 
\rule[-1ex]{0pt}{3.5ex}  \textbf{Pixel size}  & 15 $\mu$m & 18 $\mu$m  & 13 $\mu$m  \\
\hline 
\rule[-1ex]{0pt}{3.5ex}  \textbf{Pixel Scale} & 0.28"/pixel & 0.25"/pixel & 0.205"/pixel\\
\hline 
\end{tabular}
\end{center}
\end{table}



\subsection{Common Path (CP)}
\begin{figure} [ht]
\begin{center}
\begin{tabular}{c} 
\includegraphics[width=10cm]{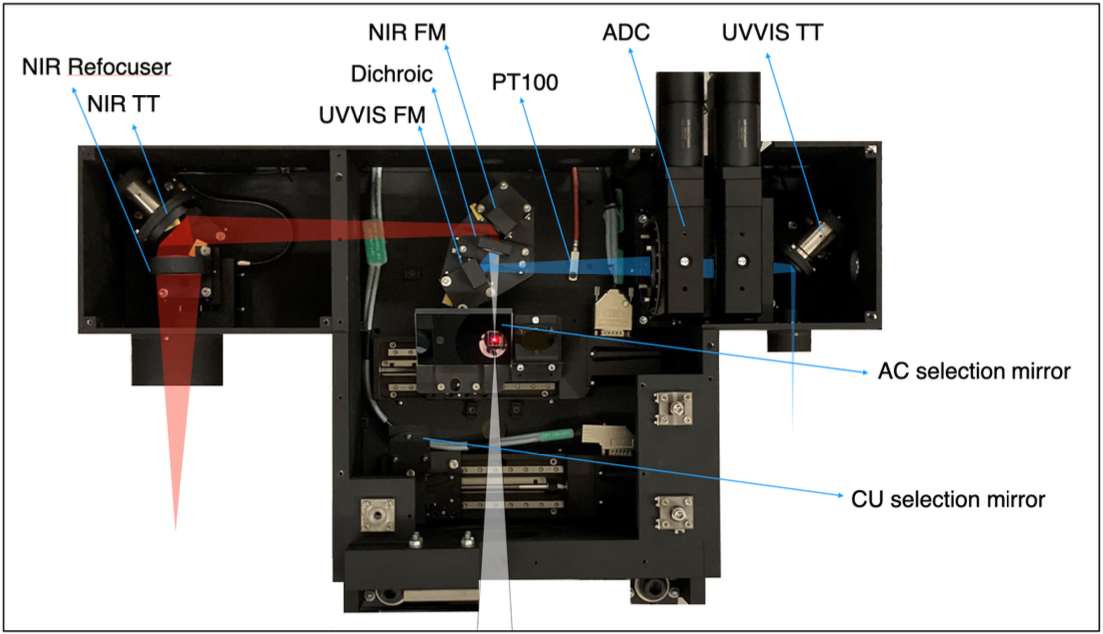}
\end{tabular}
\end{center}
\caption[example] 
   { \label{fig:SOXS-CP} SOXS Common Path, showing the components and the light path.}
\end{figure}

The common path is the heart of the SOXS\cite{kalyan-2022}. It has an optical interface with the telescope and opto-mechanical interfaces with four other SOXS sub-systems. It receives the F/11 beam from the telescope, feeds the spectrographs with an F/6.5 beam, or directs it toward the imaging camera. And, during calibration, it directs the light from the CBX to the spectrographs. Figure \ref{fig:SOXS-CP} displays the inside of the CP and also the light path. 
An ADC is correcting the atmospheric dispersion in the UVVIS light path. CP also harbors two piezo tip-tilt mirrors (marked UVVIS TT and NIR TT in Figure \ref{fig:SOXS-CP}) that can do active flexure compensation between the common path and each of the individual spectrographs. The CP's stand-alone optical quality and exit beam are measured and quantified to be within the requirements \cite{kalyan-2022}. Also, the optical efficiency of the CP is calculated to be within the requirements \cite{kalyan-2022}. The CP weighs about 47 kg.

\subsection{Calibration Unit (CBX)}
The CBX is equipped with seven lamps: the continuum lamps - QTH (from OSRAM) and Deuterium (from Hamamatsu), and arc lamps - Thorium Argon (ThAr, from Green Scientific) and four Penray lamps (Neon, Mercury, Xenon, and Argon, from Newport) \cite{hanin-2020}. For calibration purposes, the CBX produces an F/11 beam at the SOXS focal plane, which can illuminate the entire slit at the slit plane.

In addition, there is also a pinhole mode available, which will be only used for diagnostic and alignment purposes.
\begin{figure} [h]
\begin{center}
\begin{tabular}{c} 
\includegraphics[width=6.5cm]{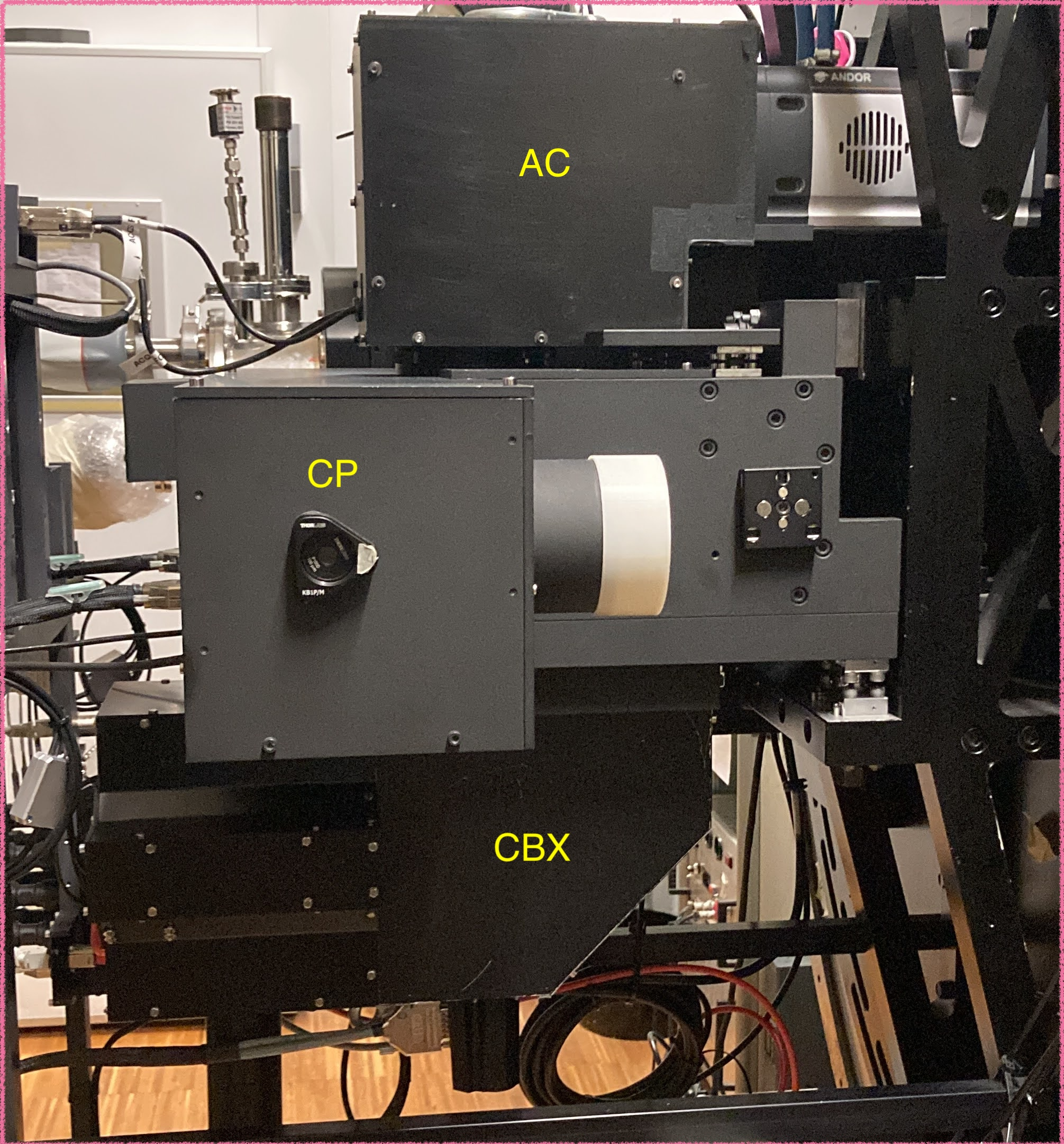}
\end{tabular}
\end{center}
\caption[example] 
   { \label{fig:SOXS-CBX-CP-AC} The calibration unit and acquisition \& imaging camera mounted to the common path, which is mounted to the SOXS flange.}
\end{figure} 
The optical box part of the CBX, holding the lamps, was modified to accommodate the flux requirements at the NIR and UVVIS detector (considering the entire optical path). The CBX weighs about 10 kg. Figure \ref{fig:SOXS-CBX-CP-AC} shows the CBX mounted to the CP.

\subsection{Acquisition and Imaging Camera (AC)}
The AC can do imaging and photometry\cite{jose2024,jose-2022}. AC will also be used during the acquisition sequence before taking spectra. The initial acquisition image will be saved and provided to the user for every spectroscopic target. While taking the spectroscopic data, part of the off-axis beam is directed toward the AC, which can be used for secondary guidance if needed.

The AC \textit{Andor} iKon‐M934  CCD is cooled thanks to a 4-stage Peltier that allows a minimum temperature of -100$^{\circ}$C. The SOXS water-cooling line removes the heat. 

The AC calibrations are performed on-sky during twilight. The AC weighs about 12 kg. Figure \ref{fig:SOXS-CBX-CP-AC} shows the AC mounted to the CP.

\subsection{UVVIS Spectrograph}

UVVIS spectrograph is conceptually and mechanically divided into two parts \cite{adam-2020}. 
\begin{figure} [h]
\begin{center}
\begin{tabular}{c} 
\includegraphics[width=10cm]{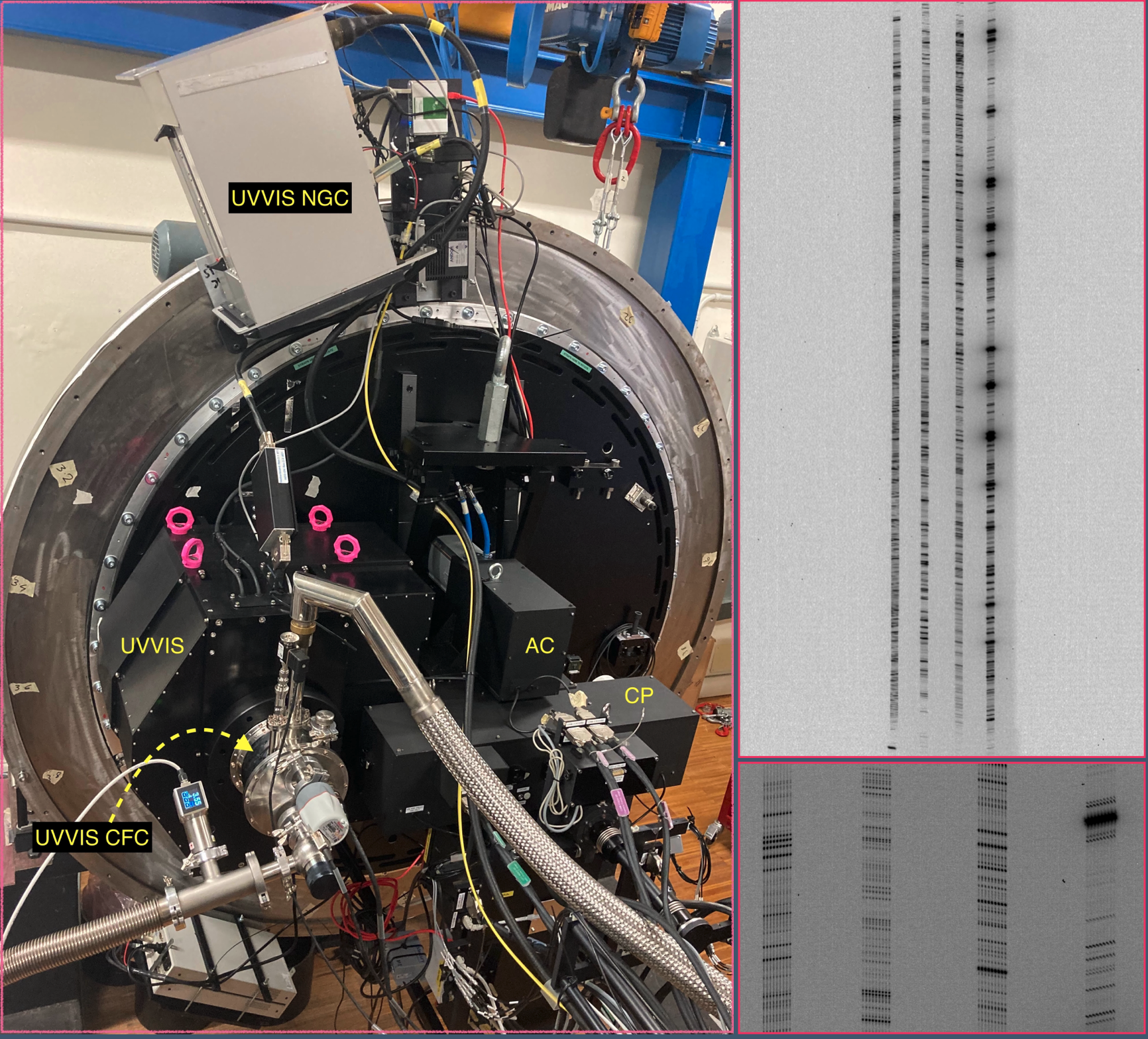}
\end{tabular}
\end{center}
\caption[example] 
   { \label{fig:SOXS-uvvis} \textit{Left}: The UVVIS spectrograph mounted to the SOXS flange. \textit{Right-top}: Raw UVVIS spectra taken with ThAr arc lamp with 1" slit and 300s exposure time. \textit{Right-bottom}: Zoomed in part of the raw UVVIS spectra taken with ThAr arc lamp with multi-pinhole and 300s exposure time. The spectral traces are r, u, g and i from left to right.}
\end{figure} 
The first part divides the CP beam into four collimated beams, each covering a different wavelength range. The second part includes the dispersers and camera, imaging the spectra of each quasi‐order to a common detector\cite{rosario2024,rosario-2022}. The dispersers are the multiple ion‐etched gratings (from the Fraunhofer Institute) optimized for the relevant beam bandwidth. The four quasi-orders are u (350 - 440 nm), g (427 - 545 nm), r (522 - 680 nm), and i (656 - 850 nm).

The UVVIS optical bench is at ambient temperature. Only the detector is cooled down using a Continuous Flow Cryostat (CFC), which uses liquid nitrogen as the coolant. 
The UVVIS integration to the SOXS is almost complete. We found residual flexures within the UVVIS system. We have found the source of the problem. The detector unit, including the CFC, is flexing due to gravity. Support structures to alleviate the problem are being made. Note that the UVVIS spectrograph weighs about 140 kg.

Figure \ref{fig:SOXS-uvvis} shows the UVVIS mounted onto the SOXS flange. Except for the flexures, UVVIS produces good spectra with spectra resolution within the requirements on static conditions. In the figure, you can see the raw spectra due to the ThAr lamp passing through a 1" slit and a multi-pinhole (for calibration).

\subsection{NIR Spectrograph}

NIR spectrograph is a fully cryogenic Echelle cross-dispersed spectrograph based on the 4C concept (Collimator Compensation of Camera Chromatism)\cite{fabrizio2024,genoni2024,fabrizio-2022,genoni-2022}. The system is compact and stiff. The NIR vessel, D-shaped, is cooled down to 145K to lower the thermal background and is equipped with a thermal filter to block any thermal radiation above 2.0 $\mu$m. The detector is cooled down to 40K. The cryogenics are operated via a closed-cycle cryocooler, with Helium gas as the coolant.
\begin{figure} [h]
\begin{center}
\begin{tabular}{c} 
\includegraphics[width=10cm]{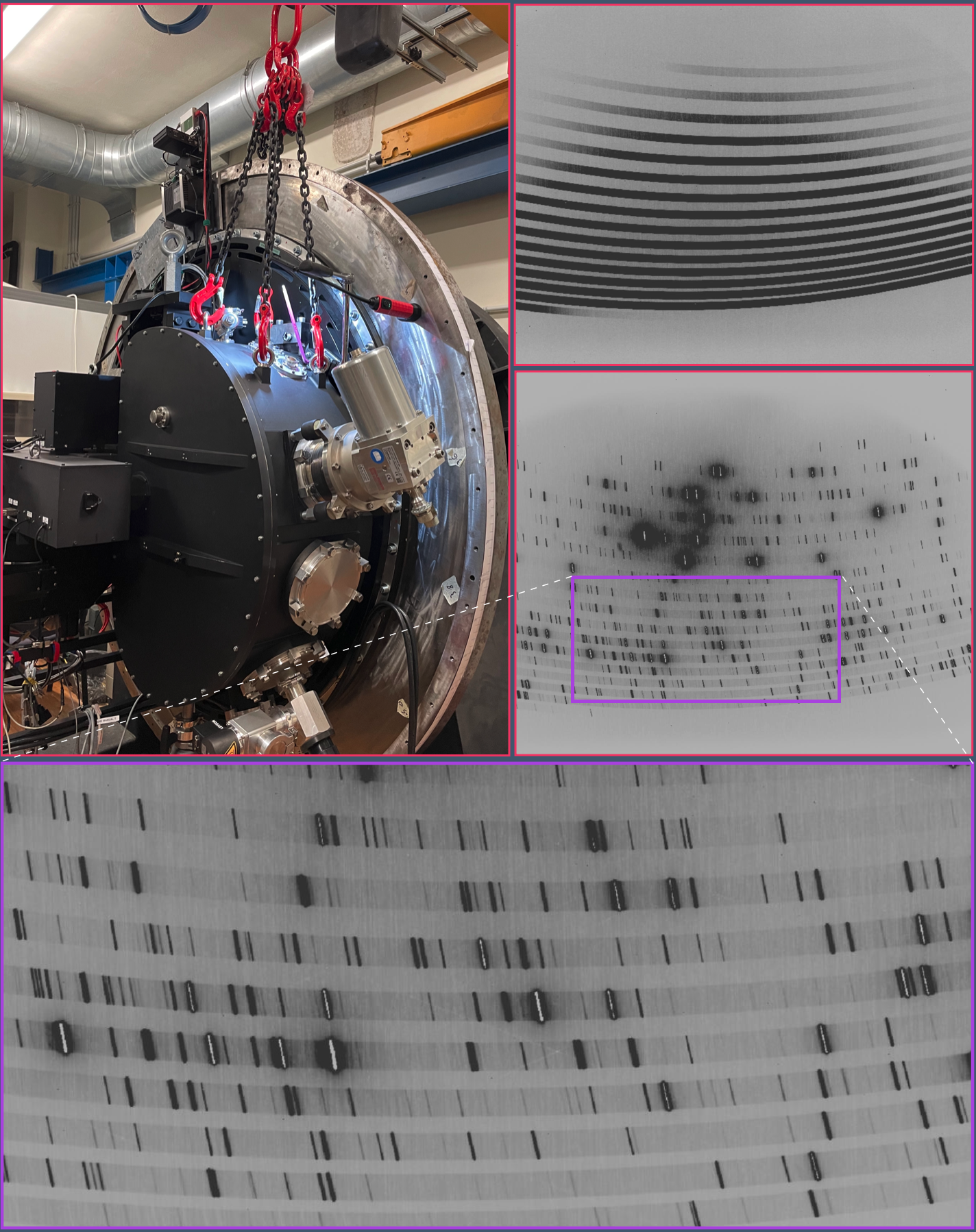}
\end{tabular}
\end{center}
\caption[example] 
   { \label{fig:SOXS-nir} \textit{Top-Left}: The NIR spectrograph being mounted to the SOXS flange. \textit{Top-Right-Top}: A raw NIR spectra taken with QTH continuum lamp with 0.5" slit and 5s exposure time. \textit{Top-Right-bottom}: A raw NIR spectra taken with arc lamps Ar, Ne, Hg, and Xe switched on with 0.5" slit and 15s exposure time. \textit{Bottom}: A zoomed in part of the arc lamp spectrum above.}
\end{figure} 

\begin{figure} [h]
\begin{center}
\begin{tabular}{c} 
\includegraphics[width=10cm]{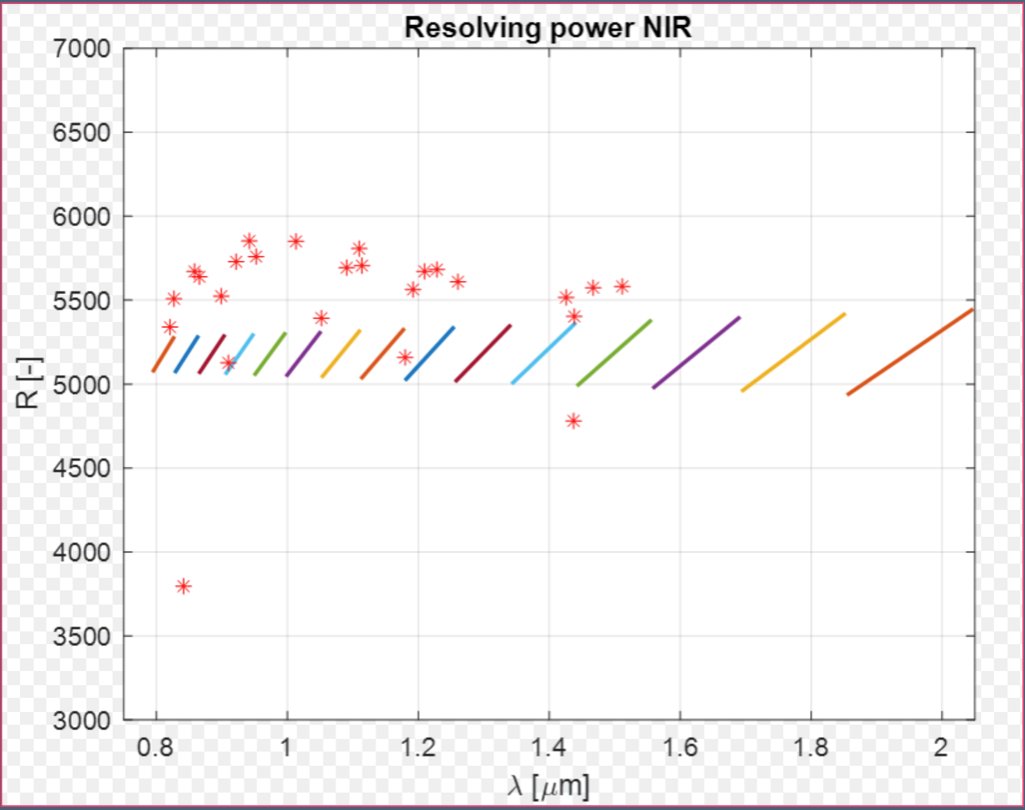}
\end{tabular}
\end{center}
\caption[example] 
   { \label{fig:nir-R} The estimated spectral resolution of R using the CBX lamp in SOXS for 1" slit.}
\end{figure}

\begin{figure} [ht]
\begin{center}
\begin{tabular}{c} 
\includegraphics[width=13cm]{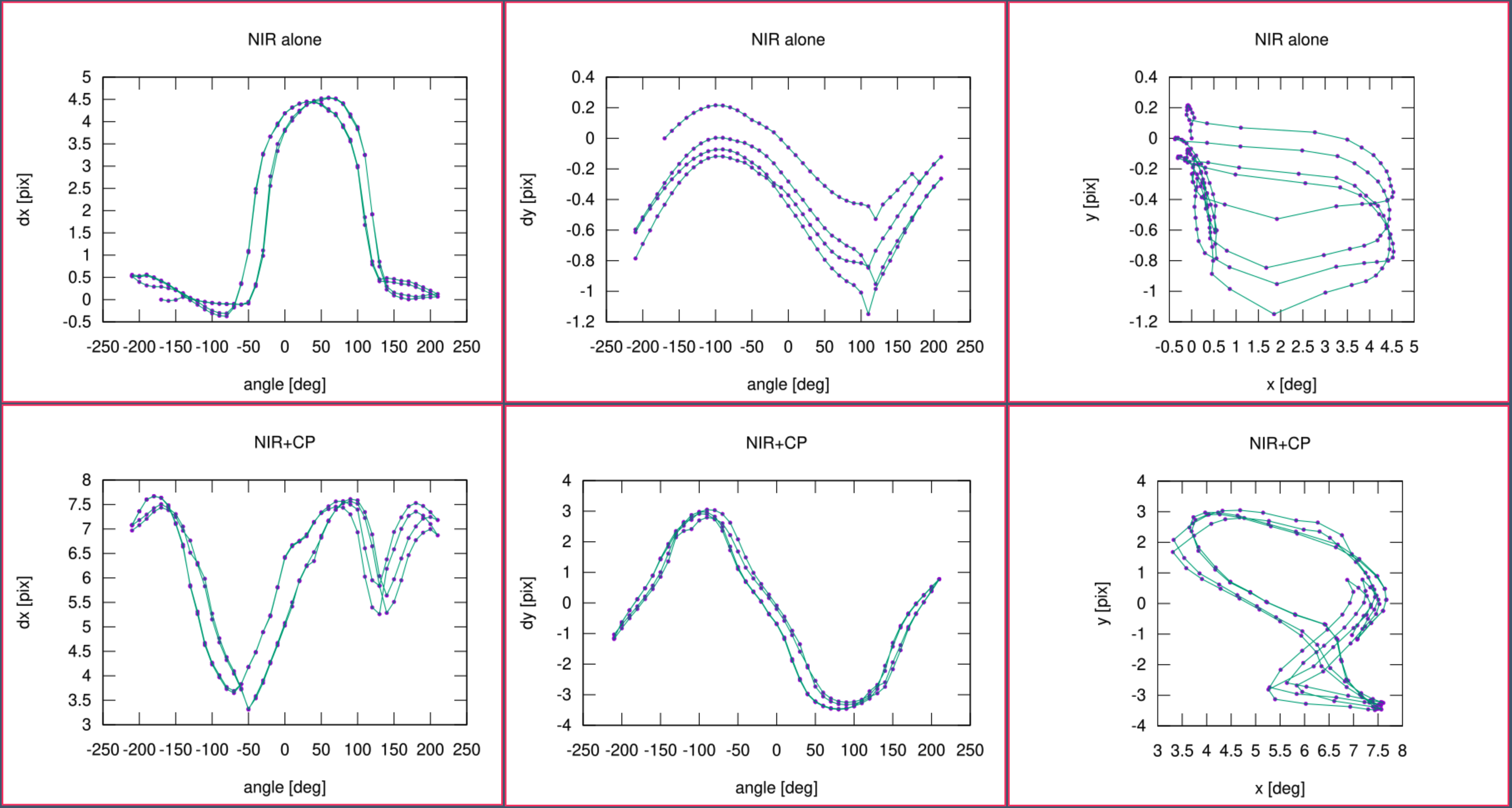}
\end{tabular}
\end{center}
\caption[example] 
   { \label{fig:nir-flex} Measured flexure of the NIR alone (top) and NIR+CP (bottom). \textit{Top-Left}: NIR alone flexure - X-direction movement. \textit{Top-Middle}: NIR alone flexure - Y-direction movement \textit{Top-Right}: NIR alone flexure in the XY plane. \textit{Bottom-Left}: NIR+CP flexure - X-direction movement. \textit{Bottom-Middle}: NIR+CP flexure - Y-direction movement \textit{Top-Right}: NIR+CP flexure in the XY plane. }
\end{figure}

After the alignment and integration to the rest of the SOXS, the optical quality of the spectra is found to be good and well within the requirements. Figure \ref{fig:nir-R} plots the measured spectral resolution of the NIR spectrograph for the entire NIR optical path starting from the calibration unit. We have measured the flexure produced by the NIR alone and between the CP and NIR (see Figure \ref{fig:nir-flex}). The optical quality is verified to be within requirements for all the derotator angles. The look-up table compensating for the flexure is ready to be tested. We will also be confirming that the flexures are repeatable and consistent. Note that the NIR spectrograph weighs about 220kg.

\subsection{Cryo-Vacuum System }
SOXS cryo-vacuum system takes care of cooling down the UVVIS CCD and the NIR vessel and its detector\cite{salvo-2022}. 

The UVVIS CCD is cooled down using a continuous flow cryostat (CFC) with liquid nitrogen as the coolant. A 120L tank provides a continuous supply of nitrogen. The nitrogen is transported from the tank via special cooling lines. These lines are designed and positioned so that they co-rotate with the SOXS corotator and, therefore, the NTT derotator. From our experience in Padova, one full tank lasts between 7 and 10 days, depending on the ambient temperature.

\begin{figure} [h]
\begin{center}
\begin{tabular}{c} 
\includegraphics[width=12cm]{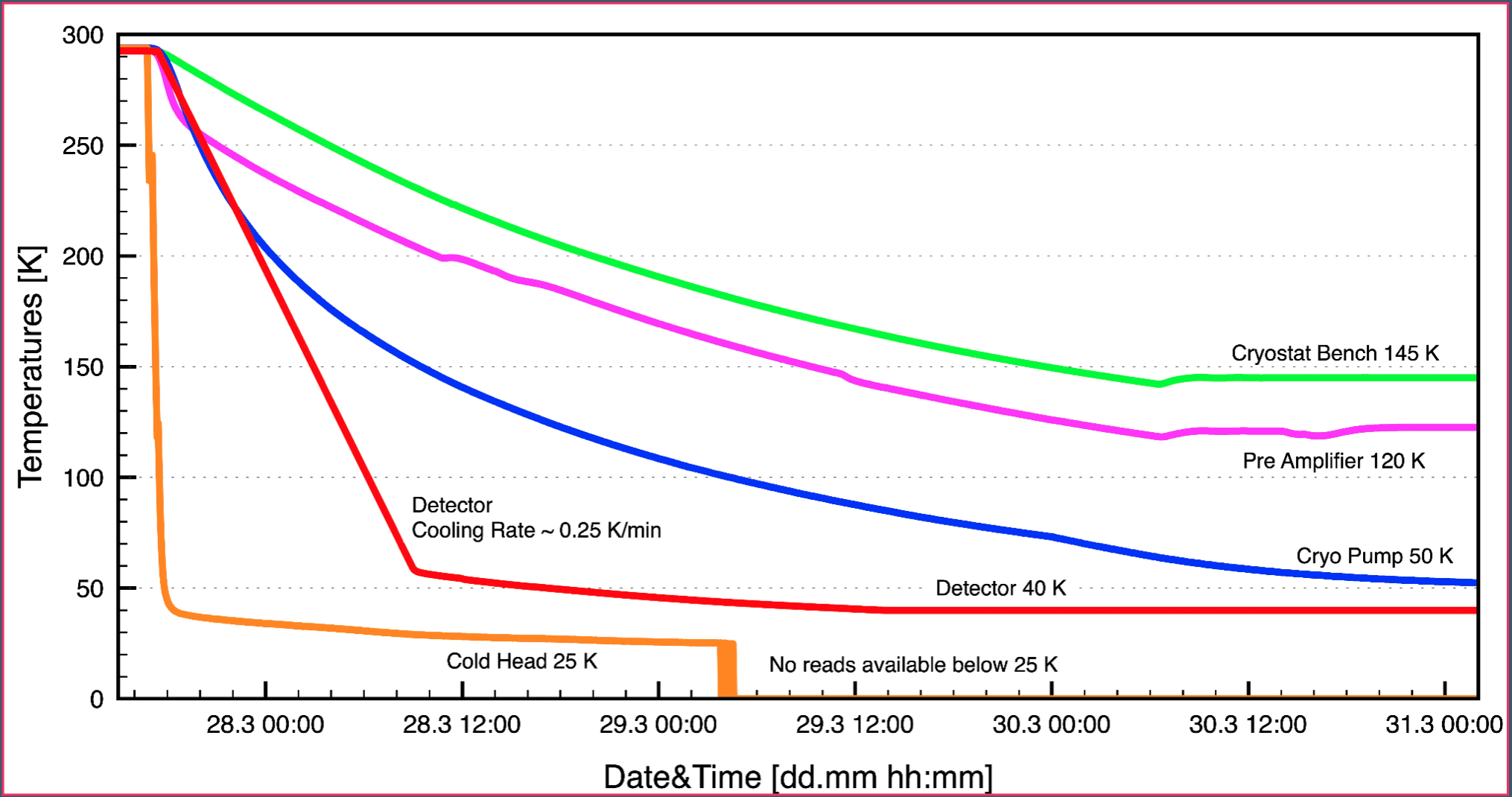}
\end{tabular}
\end{center}
\caption[example] 
   { \label{fig:nir-cryo} The cooling curves for the NIR. }
\end{figure}

NIR cryogenics is operated via a closed-cycle cryocooler, with Helium gas as the coolant. Water-cooled COOLPACK 5000i and COOLPOWER 250 MDi (both from Leybold) are the compressor and the cold-head system cooling down the NIR vessel and its detector based on the Gifford-McMahon principle. While the NIR vessel is cooled down to 145K, the NIR detector goes down to 40K. Figure \ref{fig:nir-cryo} displays the cooling curves of NIR.

The cryo-vacuum system is equipped with a pre-vacuum pump and a turbo-molecular pump to reach the necessary vacuum conditions for the cooling process. The pumps will also be used when regeneration is required. Note that when the pumps are used, the derotator position has to be fixed (or rotations are not allowed). Of course, the use of pumps is required only during daytime or regular maintenance operations. 

All the cryo-vacuum procedures are controlled by Siemens's PLC on board one of the electronics cabinets. There are dedicated, almost automatic procedures to cool down the UVVIS and NIR systems. Several alarms from the cryo-vacuum systems are connected to the SELCO alarm system. The SELCO alarm system will let the operator know if there are any potential issues with the cool-down process and, therefore, with observing using the spectrographs.

\subsection{Electronics }

The instrument control electronics (ICE) and software (ICS) control all motorized functions, sensors, and calibration lamps. The ICE is mainly based on a PLC architecture\cite{mirko2024}. The cryo-vacuum control system has an independent and separate PLC that manages and monitors all the valves, gauges, and temperature sensors needed to guarantee the proper functionality of the two spectrographs.

\begin{figure} [h]
\begin{center}
\begin{tabular}{c} 
\includegraphics[width=16cm]{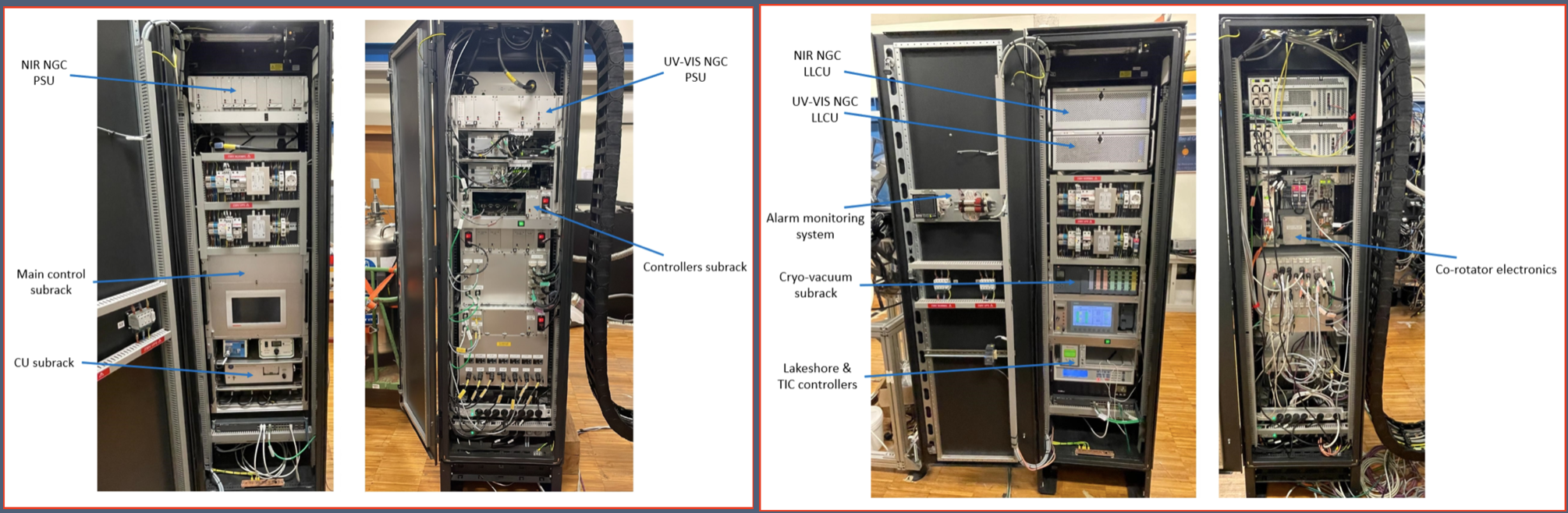}
\end{tabular}
\end{center}
\caption[example] 
   { \label{fig:electronics} \textit{Left:} Cabinet 1 front and rear views. \textit{Right:} Cabinet 2 front and rear views.
 }
\end{figure}

All the controllers and devices are installed in specific subracks hosted in two electronic cabinets. Note that the two ESO NGC controllers that control the UV-VIS and NIR detectors are installed directly on the NTT derotator. The SOXS cabinets are shown in Figure \ref{fig:electronics}.

\subsection{Instrument Control Software}
\begin{figure} [h]
\begin{center}
\begin{tabular}{c} 
\includegraphics[width=17cm]{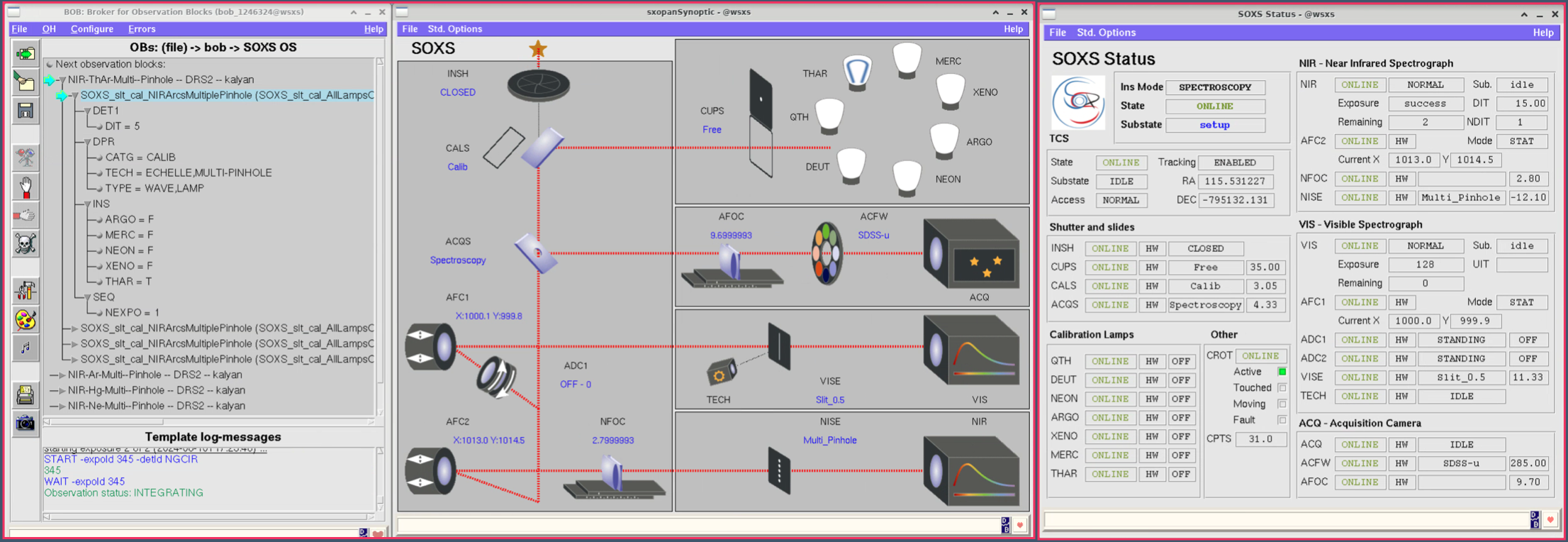}
\end{tabular}
\end{center}
\caption[example] 
   { \label{fig:ins} The BoB, Synoptic Panel, and SOXS Status GUI while running a SOXS calibration observing block. }
\end{figure}

The SOXS ICS is almost in its final stages of testing\cite{davide2024}. The AIV team has extensively tested the SOXS INS for all the features, including the templates and OBs. All the calibration templates have been tested and are working well. Figure \ref{fig:ins} displays the BOB, synoptic panel, and the SOXS status GUI while running a calibration template.

\subsection{Data Reduction Pipeline}
\begin{figure} []
\begin{center}
\begin{tabular}{c} 
\includegraphics[width=13cm]{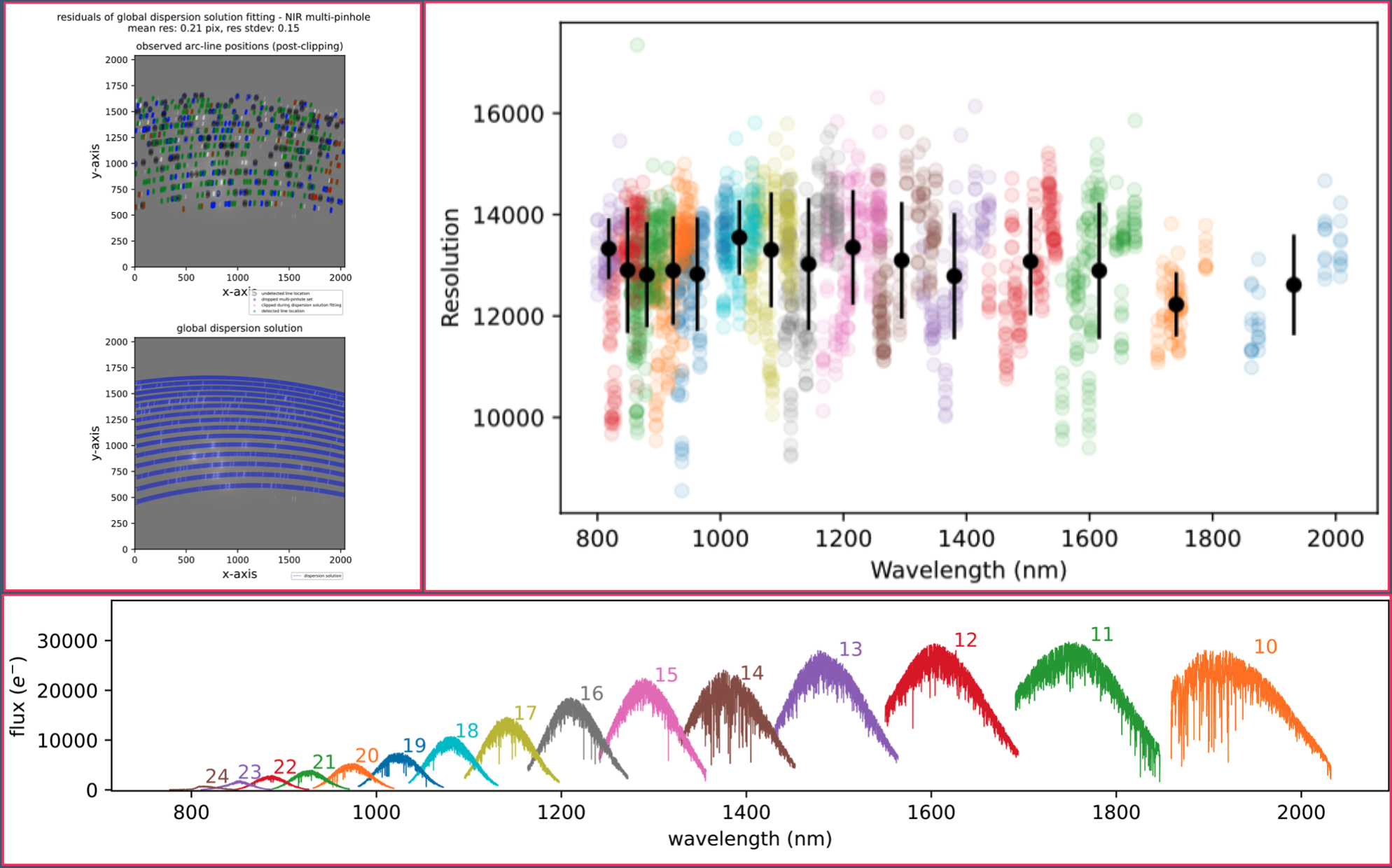}
\end{tabular}
\end{center}
\caption[example] 
   { \label{fig:drs} Plots from the output of the data reduction pipeline. \textit{Top-Left}: Quality control plot for the NIR Multi-Pinhole frame  \textit{Top-Right}: Spectral resolution estimated for a 0.5" pinhole spectra. \textit{Bottom}: All the spectral orders extracted from the NIR flat QTH calibration spectrum.}
\end{figure}

SOXS DRS is available to all the users\cite{dave-drs-2022,marco-drs-2022}. We have installed the latest version of the DRS in Padova to test the pipeline and the procedures. Additionally, the DRS team has kindly provided extra procedures useful for the AIV in Padova, which will also be used during AIV at La Silla. Figure \ref{fig:drs} shows some outputs from the DRS pipeline for the data taken using SOXS calibration templates.

\subsection{Water-cooling system}
The two electronics cabinets, the NGCs and the Andor camera, are also cooled down using water cooling lines. The AIV team has tested the cooling lines for leakage and performance. The system works as it should.

\begin{figure} [p]
\begin{center}
\begin{tabular}{c} 
\includegraphics[width=14cm]{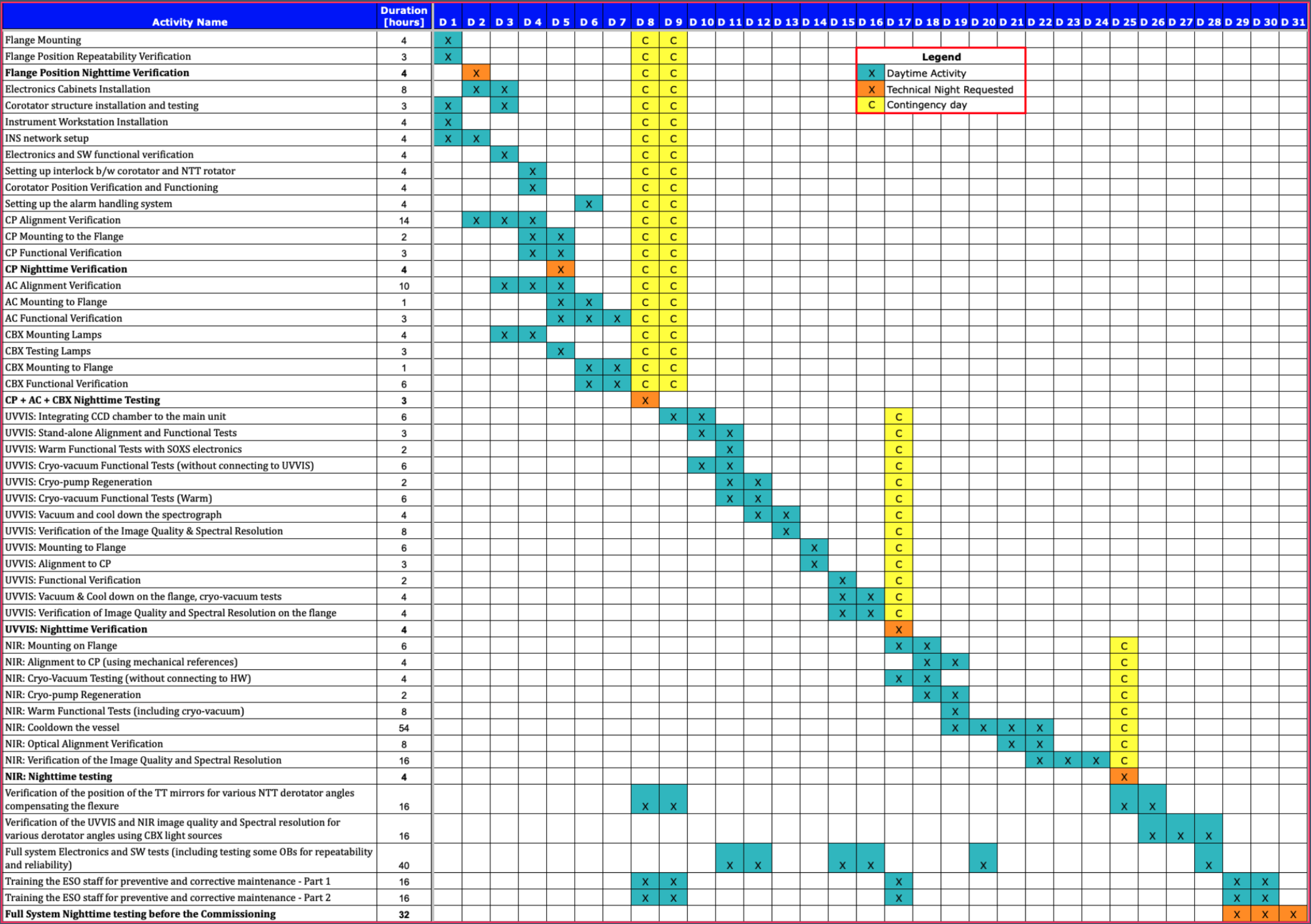}
\end{tabular}
\end{center}
\caption[example] 
   { \label{fig:aivlasilla} Tentative plan of AIV at La Silla. }
\end{figure} 
\section{What is the immediate future}
Visits related to the SOXS PAE began in June 2024. Officially, in July, we will start the PAE process. Upon successful completion of the PAE by October or early November, we will start dismounting the subsystems and components and starting the packing process. 
With the aim of shipping SOXS to La Silla by air in December 2024, we are looking forward to the AIV at La Silla in January 2025, which is the next significant milestone. The draft plan for this phase is illustrated in Figure \ref{fig:aivlasilla}.

\section{CONCLUSION}
The last two decades have witnessed many discoveries in astrophysics - gravitational wave events, gamma-ray bursts, super-luminous supernovae, accelerating universe, observations on Sagittarius A*, imaging the black hole, etc., to name a few of them. Transients is one of the astronomical research fields that has expanded significantly along with the new discoveries. With the availability of so many transient imaging surveys in the present and future (e.g., the Vera Rubin observatory), the scientific bottleneck is the spectroscopic follow-up of transients. SOXS will play a significant role in bridging this gap. It will be one of the few spectrographs on a dedicated telescope with a substantial amount of observing time to characterize astrophysical transients.
\begin{figure} [h]
\begin{center}
\begin{tabular}{c} 
\includegraphics[width=11cm]{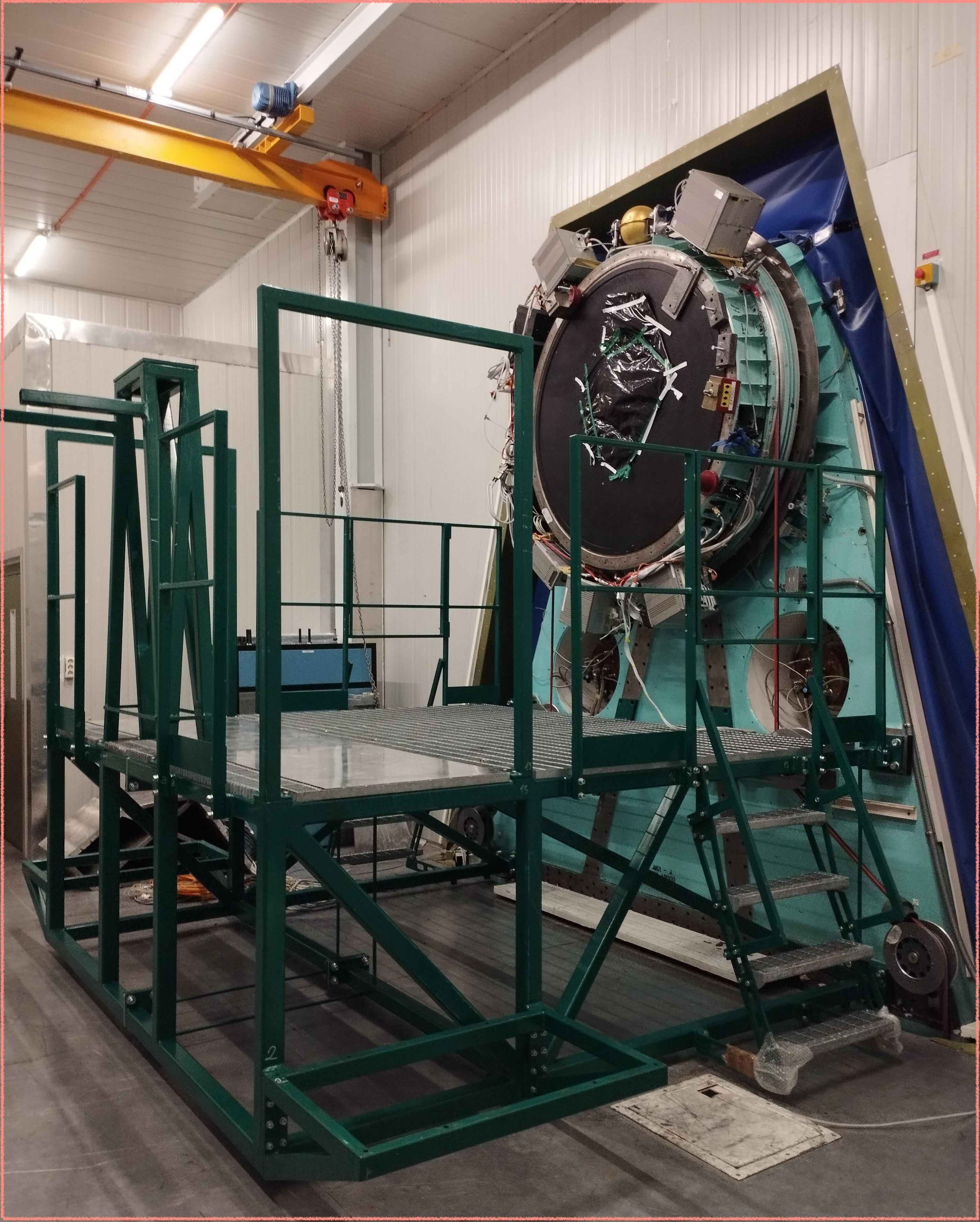}
\end{tabular}
\end{center}
\caption[example] 
   { \label{fig:platform} SOXS platform installed at La Silla. }
\end{figure} 

SOXS platform has already been installed at La Silla (see Figure \ref{fig:platform}). SOXS integration and verification are reaching the final stages. The SOXS team and the instrument are preparing for the PAE starting July 2024. Upon completing the PAE, the instrument will be dismounted, packed, and shipped to La Silla (via air), hopefully by the end of 2024. The AIV at La Silla is anticipated to be completed in January 2025, in about 30 days. Soon after, in March 2025, the nighttime commissioning will start. Hence the question: "What is your favorite transient event? SOXS is almost ready to observe!"

\acknowledgments 
The SOXS team thanks the ESO colleagues at La Silla for coordinating and supporting the installation of the SOXS platform.

KKRS and the SOXS team thank the administrative and technical staff of INAF-OAPD for all the support they received during the integration and testing of the SOXS instrument on-site.  

\bibliography{main} 
\bibliographystyle{spiebib} 

\end{document}